\documentclass[final,3p]{CSP}
\usepackage{amssymb}
\usepackage{changepage}
\usepackage{amsmath}
\usepackage{graphicx}
\usepackage{caption}
\begin{document}

\begin{frontmatter}

\title{Obstructed Atomic Limit Topological Protection in $C_4$-Symmetric Photonic Crystals for Optical Communications}

\author[charles,icmab]{Ondřej Novák}
\ead{ondrej.novak@matfyz.cuni.cz}
\author[charles]{Martin Veis}
\author[icmab]{Gervasi Herranz}

\address[charles]{Charles University, Faculty of Mathematics and Physics, Ke Karlovu 5, 121 16 Prague, Czech Republic}
\address[icmab]{Institut de Ciència de Materials de Barcelona (ICMAB-CSIC), Campus UAB, 08193 Bellaterra, Catalonia, Spain}


\begin{abstract}\rm
\begin{adjustwidth}{2cm}{2cm}{\itshape\textbf{Abstract:}} 
Recent developments in photonic topological phases have revealed that protected edge modes can emerge not only from global topological invariants, but also from symmetry-enforced polarization mismatches between distinct bulk phases. In this work, we investigate the capabilities and limitations of a square-lattice ($C_4$-symmetric) photonic crystal composed of a single dielectric material that supports interface-localized modes at the boundary between regions characterized by distinct obstructed atomic limits (OALs). These modes are confined to a common band gap and exhibit high transmission, even in the presence of structural perturbations. 

Our analysis reveals that the interface modes are stabilized by a mismatch in the position of Wannier centers between the two adjoining crystals. We demonstrate nearly lossless transmission around sharp turns and through localized defects, though the robustness depends asymmetrically on the side of perturbation, reflecting the partial nature of the protection. We also show that increasing the number of photonic crystal periods surrounding the interface enhances both modal confinement and spectral stability. These findings establish polarization mismatch between OALs as a practical and fabrication-compatible mechanism for engineering robust photonic transport in \(C_4\)-symmetric systems.
\end{adjustwidth}
\end{abstract}
\end{frontmatter}

\section{Introduction}
\noindent
\label{intro_sec}


Over the past decade, the field of topological protection in photonics has witnessed significant growth and sustained interest~\cite{bliokh2015quantum,funayama2022control,smirnova2020nonlinear,wang2022short,tang2022topological,lu2014topological,wang2020topological,khanikaev2017two,zhirihin2021topological,ozawa2019topological}. The most promising applications of topologically robust states of light are emerging in optical insulators~\cite{tkachov2015topological,jalas2013and}, waveguides and couplers~\cite{kagami2020topological,kagami2021highly,jin2022manipulation,kang2023topological,du2022optimal,yang2022observation,iwamoto2021recent,hauff2022chiral}, topological lasers~\cite{gong2020topological}, on-chip nanobeam filters~\cite{liu2024chip}, wavelength-division multiplexing~\cite{he2024wavelength}, and logic gate architectures~\cite{zhang2023experimental}. Fabrication~\cite{rechtsman2023reciprocal} and characterization techniques~\cite{kang2024multiband,liu2020z2,arora2022breakdown,li2021experimental,yoshimi2021experimental,shalaev2019optically,wu2024spin,skirlo2015experimental} are now well-established, with growing emphasis on integration with scalable photonic platforms~\cite{kagami2020topological}.

The suppression of backscattering in symmetry-protected edge states is particularly valuable to ensure robust light transmission even in the presence of structural imperfections~\cite{xu2016accidental,blanco2020tutorial,devescovi2024tutorial,huang2022topological,wu2015scheme,xu2023absence,liu2018topological,wang2008reflection}. Despite extensive work on established lattice geometries, exploring alternative photonic crystal (PC) symmetries remains essential for applications such as integrated logic, where geometry dictates routing constraints and density~\cite{he2022topological}. In the case of the $C_4$ lattice, realizations of topological phases have so far been limited to Chern and valley-Chern phases~\cite{tang2022topological,iwamoto2021recent}. However, Chern insulators generally require magnetic or gyromagnetic effects, limiting them to the microwave regime due to material constraints~\cite{newnham2005properties,tsutaoka2011permeability}, while valley-Chern phases are inherently limited to part of the Brillouin zone.

A promising alternative is offered by the concept of obstructed atomic limits (OALs). In these systems, although the bands are topologically trivial in the conventional sense (i.e., they admit localized Wannier representations), the Wannier centers are quantized at non-atomic positions, such as cell edges or corners \cite{vaidya2023topological}. This configuration leads to a quantized bulk polarization and symmetry-protected interface states when two phases with different Wannier center positions are juxtaposed~\cite{wang2022short,peterson2018quantized}. Unlike Chern or $\mathbb{Z}_2$ topological phases, OALs do not require time-reversal symmetry breaking or spin-like degrees of freedom and are thus naturally suited to all-dielectric photonic platforms.

The $C_4$ lattice is particularly attractive for realizing such behavior due to its compatibility with orthogonal logic architectures~\cite{sharma2021common}. Orthogonal layouts facilitate high-density photonic integration by simplifying routing and reducing parasitic effects. Moreover, the high symmetry of the $C_4$ lattice enables compact and fabrication-friendly designs for logic elements and other functional devices, making it ideal for integrated topological photonics.

In this paper, we present a $C_4$-symmetric photonic crystal design that exhibits a polarization contrast between two structural variants. Using Wilson loop analysis, we identify an obstructed atomic limit in one of the structures, characterized by Wannier centers pinned to the boundaries of the unit cell rather than its center. This mismatch between the two regions gives rise to symmetry-protected interface modes, offering directional and partially robust light transmission. Unlike previous designs that require complex geometries~\cite{zhang2024topological} or multi-material platforms~\cite{xiong2022topological}, our implementation is based on a single-material InP photonic crystal, designed for operation at 1550\,nm, a wavelength relevant for telecommunications. Our approach combines conceptual simplicity with practical compatibility with modern photonic integration technologies.

\subsection{Photonic eigenproblem}
\label{mathematics}
\noindent
Topological invariants of photonic systems can be derived from Maxwell's equations. In two dimensions, these equations decouple into transverse electric (TE) and transverse magnetic (TM) modes \cite{jones2013theory}. Here, we focus on TM modes, allowing us to formulate a wave equation for the \( z \)-component of the electric field:

\begin{equation}
\nabla \times \nabla \times \mathbf{E}_z (\mathbf{r}) = \left( \frac{\omega}{c} \right)^2 \epsilon(\mathbf{r}) \mathbf{E}_z (\mathbf{r}),
\end{equation}

where \(\epsilon(\mathbf{r})\) represents a spatially and possibly frequency-dependent dielectric tensor. This wave equation can be reframed as an eigenvalue problem:

\begin{equation}
\hat{A} E_z (\mathbf{r}) = a E_z (\mathbf{r}),
\end{equation}

where \(\hat{A}\) is an operator acting on \(E_z\), with eigenvalue \(a\). These eigenvalues form energy bands, and in general, \(\hat{A}\) has an infinite number of eigenvalues; however, we focus only on the lowest few. We express eigenvalues in reduced units, \(\frac{\omega a}{2 \pi c}\), equivalent to \(\frac{a}{\lambda}\), where \(a\) denotes the lattice parameter. Given the periodicity of \(\epsilon(\mathbf{r})\), Bloch's theorem allows us to isolate the periodic component of \(E_z\) \cite{blanco2020tutorial, watanabe2019proof}.

\begin{equation}
    E_{z,k} (\mathbf{r}) = e^{i \mathbf{k} \cdot \mathbf{r}} u_k (\mathbf{r}),
\end{equation}

where $\mathbf{k}$ is a reciprocal vector.\\

The primary focus of this paper is the function \( u_k (\mathbf{r}) \), which we obtain using the MPB software package \cite{johnson2001block}. \( u_k (\mathbf{r}) \) is a complex function, so to compare the fields and visualize them, we use the real part with a fixed phase option provided by MPB. Alternative solvers, such as COMSOL or FDTD methods \cite{oskooi2010meep,inan2011numerical,sullivan2013electromagnetic}, could also be used to achieve similar results. For further details on MPB calculations, please refer to \ref{appMPB}. Although \( u_k (\mathbf{r}) \) technically represents only the periodic component of the eigenmodes, we will refer to \( u_k (\mathbf{r}) \) as the eigenmode or the Bloch function throughout this paper.

\subsection{Eigenfunctions}
\label{eigenfunctions}
\noindent
In the context of OALs, the key topological feature is the quantized position of Wannier centers within the unit cell, rather than the presence of spin-like degrees of freedom or time-reversal symmetry \cite{proctor2020robustness}. While OAL systems are topologically trivial in the sense that they admit localized Wannier functions, also referred to as "Wannierizable", their polarization can be nonzero and quantized, leading to topologically distinct crystalline phases \cite{vaidya2023topological}. Therefore, an interface between two such phases with different polarization supports symmetry-protected interface-localized modes.

In our design process, we use band-symmetry inversion at the $\Gamma$ point as a diagnostic tool to locate potentially nontrivial phases. Specifically, we monitor the reordering of photonic bands associated with distinct orbital-like eigenmode symmetries, such as $p$ and $d$ modes. This reordering often signals a change in the real-space location of Wannier centers, central to OAL topology. However, this inversion alone does not constitute a topological invariant; instead, we rely on Wilson loop calculations to confirm the presence of a quantized polarization shift \cite{vaidya2023topological}.

To implement our strategy we use the fact that the eigenmodes at the $\Gamma$ point exhibit symmetries analogous to atomic orbitals, which allows classification into modal types such as $s$, $p$, and $d$. We observe that $p$-type states appear in two symmetry-distinct forms, namely, the conventional $p_x$ and $p_y$ orbitals, and the symmetry-adapted linear combinations $p_x \pm p_y$. The $d$-type states appear as $d_{xy}$ and $d_{x^2 - y^2}$. The structural transition we design swaps the ordering of the $p$ and $d$ states at the $\Gamma$ point, suggesting a change in the underlying Wannier center configuration.

\subsection{Wilson loop calculations}
\label{wilsonloop}
\noindent
To probe the topological properties of the bulk bands, particularly their polarization and obstructed atomic character, we compute the Wilson loop (also referred to as the Wannier center flow) for selected sets of eigenmodes. Starting from the periodic Bloch functions \( u_{n\mathbf{k}}(\mathbf{r}) \) introduced in Eq.~(3), the Wilson loop operator is constructed as a path-ordered product of overlap matrices along a discrete, closed path in reciprocal space \cite{blanco2020tutorial}. Specifically, we compute it along the \( k_x \)-direction at fixed \( k_y \), forming a closed loop across the Brillouin zone:

\begin{equation}
\mathcal{W}(k_y) = \lim_{N \to \infty} \prod_{j=0}^{N-1} F_{j,j+1}, \quad \text{where} \quad [F_{j,j+1}]_{mn} = \langle u_{m,\mathbf{k}_j} | u_{n,\mathbf{k}_{j+1}} \rangle.
\end{equation}

Here, \( m,n \) index the selected bands, and the product is path-ordered in increasing \( k_x \). The overlap matrices \( F_{j,j+1} \) capture the parallel transport of Bloch states between neighboring \( \mathbf{k} \)-points. The Wilson loop operator \( \mathcal{W}(k_y) \) has eigenvalues of unit magnitude, \( \lambda_i(k_y) = e^{i\theta_i(k_y)} \), with the phases \( \theta_i(k_y) \in [-\pi, \pi) \) representing the average positions of the hybrid Wannier centers along the real-space \( x \)-direction. Since the Wilson loop is computed over \( k_x \), these phases provide a band-resolved measure of the bulk polarization along \( x \).

A constant set of \( \theta_i(k_y) \) across the Brillouin zone indicates vanishing polarization, while a displaced or winding Wilson loop signals a topologically distinct phase. In particular, a shift in the average Wannier center between two crystal structures signifies a transition between a trivial and an OAL phase \cite{proctor2020robustness}. The presence of localized interface states at the boundary of such phases is then a direct consequence of bulk-boundary correspondence. Our Wilson loop computations are performed using the discretized parallel transport method, following the procedure described in Ref.~\cite{blanco2020tutorial}. The Bloch functions were obtained from MPB simulations. 

\section{Design space and symmetry-guided band engineering}
\label{sec:seed_structure}

\begin{figure}[t]
\centering
\includegraphics[width=8cm]{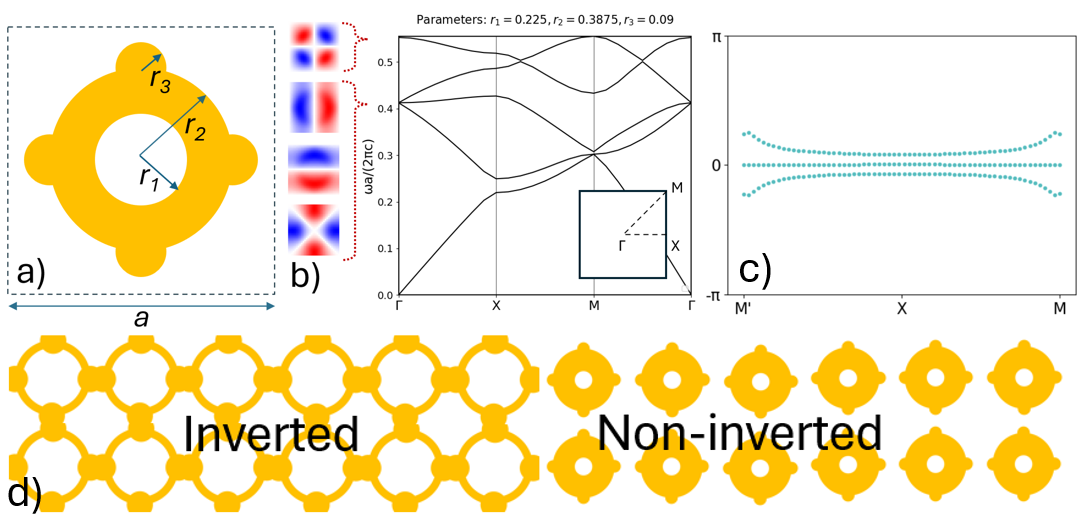}
\caption{\textbf{Seed structure and polarization-guided design.} (a) Geometry of the photonic crystal unit cell composed of InP (orange) in air. The design is defined by three tunable parameters: $r_1$ (inner radius), $r_2$ (outer radius), and $r_3$ (lobe radius), with the outer lobes positioned symmetrically on the ring to preserve $C_4$ symmetry. (b) Electric field profiles and band structure of a high-symmetry seed configuration exhibiting a triple degeneracy at the $\Gamma$ point, involving two $p$-like and one $d$-like mode. (c) Wilson loop of the seed configuration calculated over the three lowest bands, showing smooth eigenphase evolution and non-quantized polarization. (d) Comparison of two crystal configurations derived by tuning geometric parameters: an inverted structure with displaced Wannier centers and a non-inverted structure with centered polarization. These serve as the basis for constructing the topological interface studied in this work.}
\label{fig:seedstructure}
\end{figure}

We begin by introducing the photonic crystal unit cell shown in Fig.~\ref{fig:seedstructure}(a), consisting of an indium phosphide (InP) ring with four lobes arranged in a square lattice with $C_4$ symmetry. This structure is parameterized by three radii: $r_1$ (inner hole), $r_2$ (outer ring), and $r_3$ (lobe size). These parameters allow for systematic tuning of the band structure while preserving lattice symmetry. Our initial goal was to explore the possibility of achieving $\mathbb{Z}_2$-like topological protection via band inversion, inspired by analogies with honeycomb lattices and Dirac-cone physics \cite{tkachov2015topological,xu2016accidental,huang2022topological}. Although a true double Dirac cone configuration could not be found, we identified a symmetric geometry that exhibits a triple degeneracy at the $\Gamma$ point, as shown in the band structure in Fig.~\ref{fig:seedstructure}(b). This degeneracy involves two $p$-like modes and one $d$-like mode, whose real-space field profiles are shown alongside the band structure. We refer to this configuration as a seed structure, marking a high-symmetry point in parameter space near a transition between topologically distinct crystalline phases.

By tuning the geometry away from this degeneracy point, we obtain two configurations: one where the $d$-like mode lies below the $p$-like modes (inverted), and one with the reverse ordering (non-inverted), as shown in Fig.~\ref{fig:seedstructure}(d). The band reordering correlates with a shift in the position of Wannier centers, a defining feature of obstructed atomic limits. To validate this, we compute the Wilson loop for the three lowest bands of the seed configuration, shown in Fig.~\ref{fig:seedstructure}(c). The smooth, non-quantized flow of the Wilson loop phases confirms that this configuration lies near a transition point but does not yet exhibit topological polarization. By contrast, the inverted and non-inverted structures yield Wilson loops with distinct center positions, as presented in later sections. As discussed previously, band inversion is not a topological invariant but serves as a practical design guideline for identifying parameter regimes where bulk polarization transitions may arise. Therefore, the seed structure acts as a reference for navigating the configuration space and pinpointing obstructed atomic limit phases that support symmetry-protected interface states.

\section{Results}
\label{results}

\begin{figure*}[t]
\centering
\includegraphics[width=\textwidth]{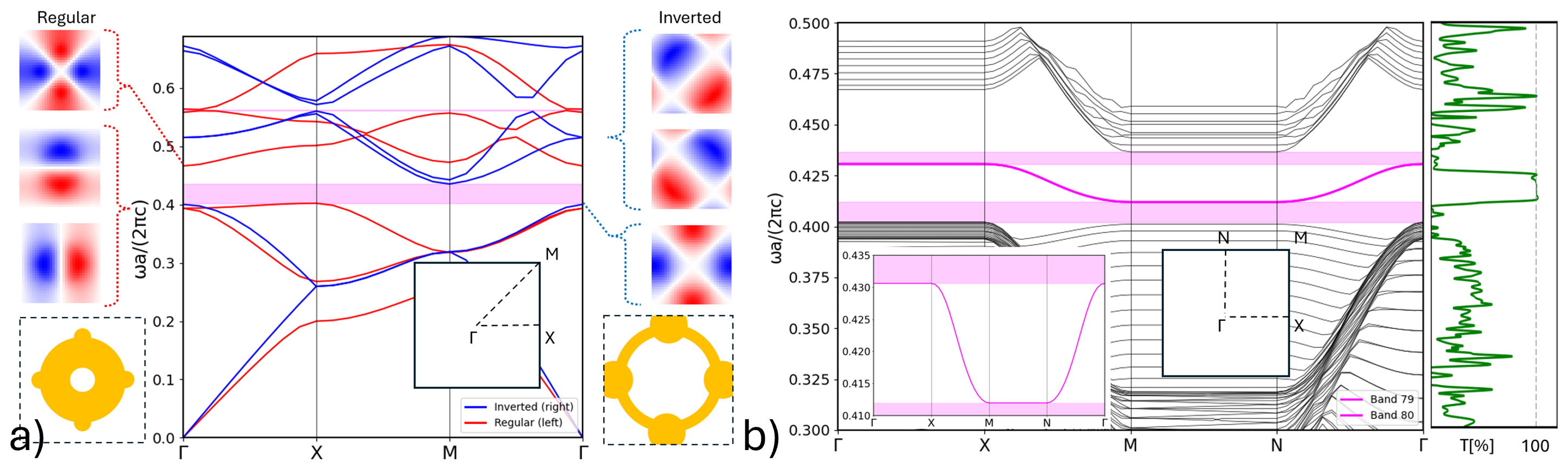}
\caption{a) \label{reduced_bandgap} Photonic band-structures obtained for crystals designed according to elementary cells defined in Figure \ref{fig:seedstructure}a. The regular (non-inverted band structure, solid red line) crystal is defined according to parameters: $r_1=0.105a$, $r_2=0.35a$, $r_3=0.072a$, while the band-inverted photonic crystal (solid blue line) is defined by parameters $r_1=0.345a$, $r_2=0.425a$, $r_3=0.108a$. The band structures of both crystals are overlaid to emphasize the overlapping band-gap and the ordering of eigenfunctions shown as the red and blue field distributions around the figure (see section \ref{mathematics}). The broad pink strip represents the band-gap common for both designs. b) Band-structure of a super-cell consisting of sixteen horizontally stacked regular cells and sixteen band-inverted cells. The broad-pink strip represents the original band-gap from \ref{reduced_bandgap}a. In this band-gap, the solid magenta lines correspond to the topological modes. The green line is a spectral dependence of transmission calculated by FDTD method.}\label{reduced_bandstructure}
\end{figure*}

\noindent
We center our investigation on a photonic topological interface formed between two crystalline phases: one with a regular, non-inverted band structure, and one with a symmetry-inverted configuration. To identify suitable configurations, we explored the geometric parameter space ($r_1$, $r_2$, $r_3$) to locate regions where the photonic crystal exhibits either regular or inverted modal ordering. Figure~\ref{reduced_bandstructure}a shows the band structures of both crystal types overlaid in the same plot. The regular, non-inverted crystal is shown in red and defined by parameters $r_1 = 0.105a$, $r_2 = 0.35a$, $r_3 = 0.072a$, while the inverted configuration (blue) corresponds to $r_1 = 0.345a$, $r_2 = 0.425a$, $r_3 = 0.108a$. The key distinction lies in the reordering of $p$- and $d$-like eigenmodes at the $\Gamma$ point, which serves as an initial indicator of topological polarization contrast.

\begin{figure}[t]
\centering
\includegraphics[width=\textwidth]{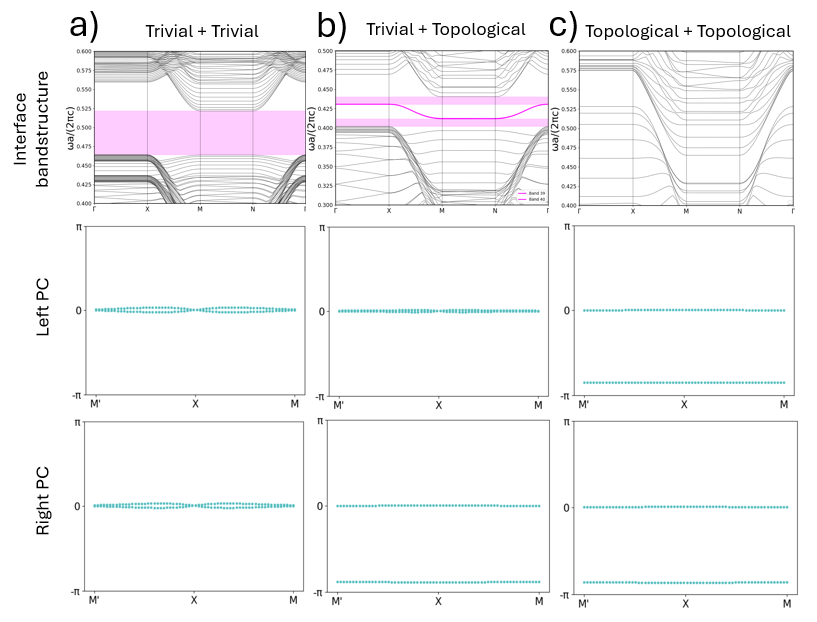}
\caption{
\textbf{Interface modes and Wilson loop analysis for three photonic crystal combinations.}
Each column represents one interface scenario created by joining two photonic crystals with distinct unit cell parameters denoted as Left PC and Right PC. 
Top row: Interface band structures computed using a supercell geometry. Gap-localized modes are visible only in case (b).
Middle and bottom rows: Wilson loop spectra for the left and right photonic crystals forming each interface, respectively. 
Case (a) shows an interface between two trivial configurations with matching Wilson loops and no interface modes. 
Case (b) corresponds to a mismatch between a trivial and an obstructed atomic limit phase, confirmed by a Wilson loop shift; this results in a pair of interface-localized modes within the bandgap. 
Case (c) involves two topological (OAL) phases with identical Wannier center positions and thus no protected interface modes. 
This analysis confirms that the presence of localized interface states arises from a polarization mismatch between the bulk phases.
}
\label{fig:wilson_bandcases}
\end{figure}

To verify the topological character of these configurations, we computed the Wilson loop spectra for both crystals, shown in Fig.~\ref{fig:wilson_bandcases}. The trivial configuration exhibits flat Wilson loop phases centered at zero, indicating vanishing polarization. In contrast, the inverted structure shows a clear shift in the average phase, reflecting a nontrivial displacement of the Wannier centers. This quantized mismatch in bulk polarization confirms that the two crystals belong to distinct OAL phases, and thus can support symmetry-protected interface states.

To test this, we constructed a supercell composed of 16 unit cells of the regular crystal interfaced with 16 unit cells of the inverted one. The resulting band structure, shown in Fig.~\ref{reduced_bandstructure}b, reveals two interface-localized modes (magenta lines) traversing the common bandgap. These modes are well confined and exhibit linear dispersion. The corresponding transmission spectrum, computed using FDTD (green line), shows near-unity transmission at the interface mode frequencies, with negligible transmission elsewhere due to the photonic bandgap. Note that the slight noise in the transmission spectrum arises from the Fourier transform of a finite-duration time pulse used in the FDTD simulation. These effects are well understood and discussed further in \ref{appFDTD}. For design purposes, we interpret the spectral features qualitatively and confirm that high transmission is isolated to the gap-localized modes.

Altogether, these results confirm that the presence of the edge-localized interface modes is due to a polarization mismatch between two crystalline phases in distinct obstructed atomic limits. Unlike full $\mathbb{Z}_2$ topological protection, this form of protection is symmetry-protected and is robust as long as the interface preserves the underlying lattice symmetry. In the following sections, we investigate the extent to which this protection is retained under structural perturbations and interface defects.

\subsection{Mode robustness}
\label{sec:defect_robustness}

\noindent
Although the topological modes are confined within a photonic band gap, thus preventing coupling to bulk propagating states, certain types of local perturbations can still affect their transmission. In contrast to fully $\mathbb{Z}_2$-protected systems, where sharp turns and local defects do not disrupt propagation, our simulations reveal an asymmetry in how the mode responds to changes placed on the left versus right side of the interface. This behavior reflects the partial nature of topological protection arising from a bulk polarization mismatch, rather than from a nontrivial $\mathbb{Z}_2$ invariant.

\begin{figure*}[t]
\centering
\includegraphics[width=\textwidth]{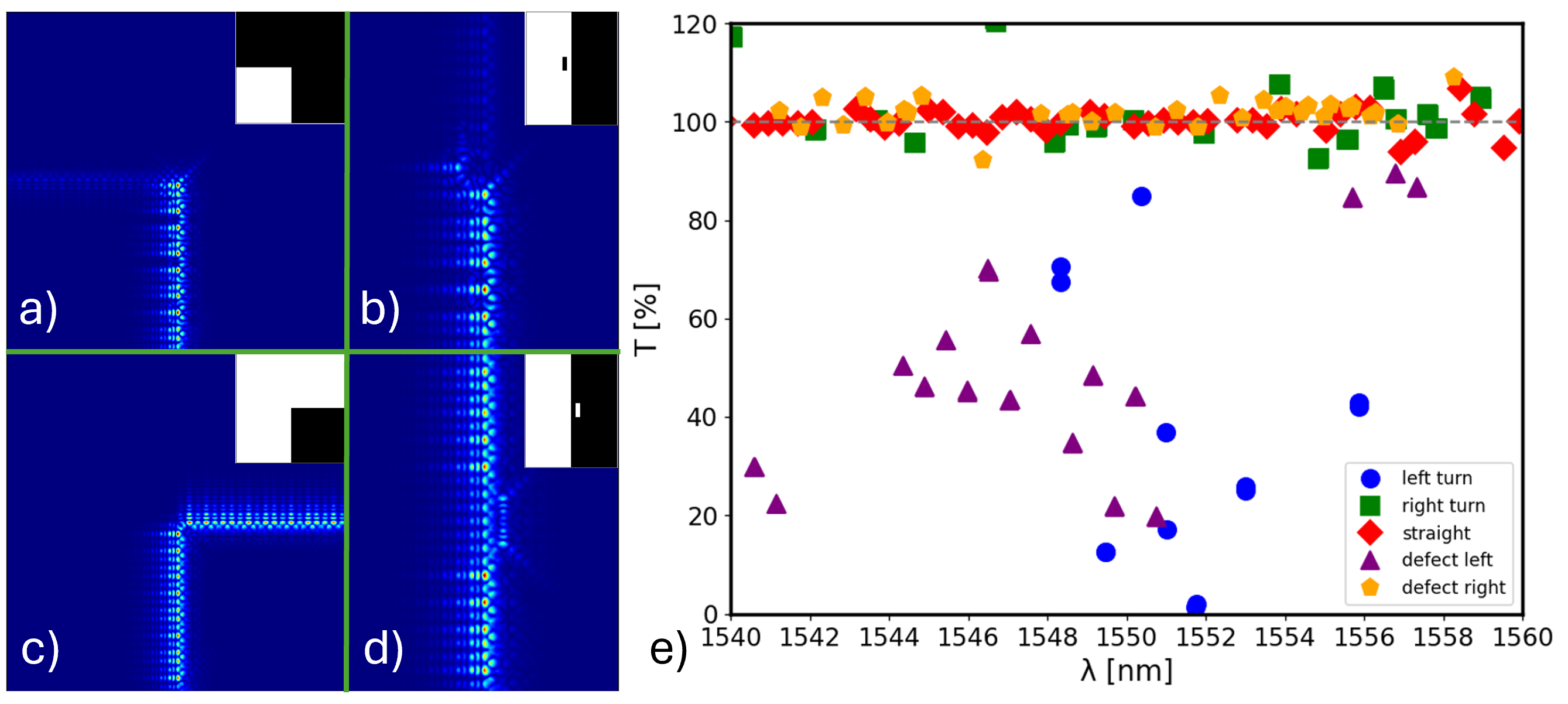}
\caption{
\textbf{Transmission robustness under structural perturbations.}
(a, c) $E_z$ field profiles for left and right 90$^\circ$ turns along the topological interface. 
(b, d) Field profiles for defects introduced one cell away from the interface, on the left (b) and right (d) side respectively. 
White regions represent trivial cells, black indicate inverted cells. 
(e) Transmission spectra for each configuration, showing strong transmission asymmetry: right-side changes preserve high transmission, while left-side changes lead to partial suppression. 
}
\label{transmission_maps}
\end{figure*}

We evaluated the robustness of the interface mode in two complementary sets of simulations using the finite-difference time-domain (FDTD) method. All simulations were performed using the Lumerical solver with a spatial resolution of 54 mesh points per unit cell, sufficient for convergence. In the first set of simulations, the interface was routed into a left or right 90° turn, with field profiles shown in Fig.~\ref{transmission_maps}a and \ref{transmission_maps}c. We also introduced a structural defect near the interface: a column of three unit cells was swapped from inverted to non-inverted (or vice versa), placed one unit cell away from the interface on either the left or right side (Fig.~\ref{transmission_maps}b,d). In all cases, the white regions represent regular (trivial) unit cells and black regions denote inverted (topological) ones. The transmission spectra for each configuration are shown in Fig.~\ref{transmission_maps}e. Right-side modifications, including bends and defects, consistently maintain near-100\% transmission, while left-side perturbations result in strong scattering and reduced transmission. This asymmetry suggests that the edge mode is not strictly protected against all symmetry-breaking perturbations, but instead relies on partial symmetry conditions tied to the local lattice environment and the real-space geometry of the Wannier centers.

\begin{figure*}[t]
\centering
\includegraphics[width=\textwidth]{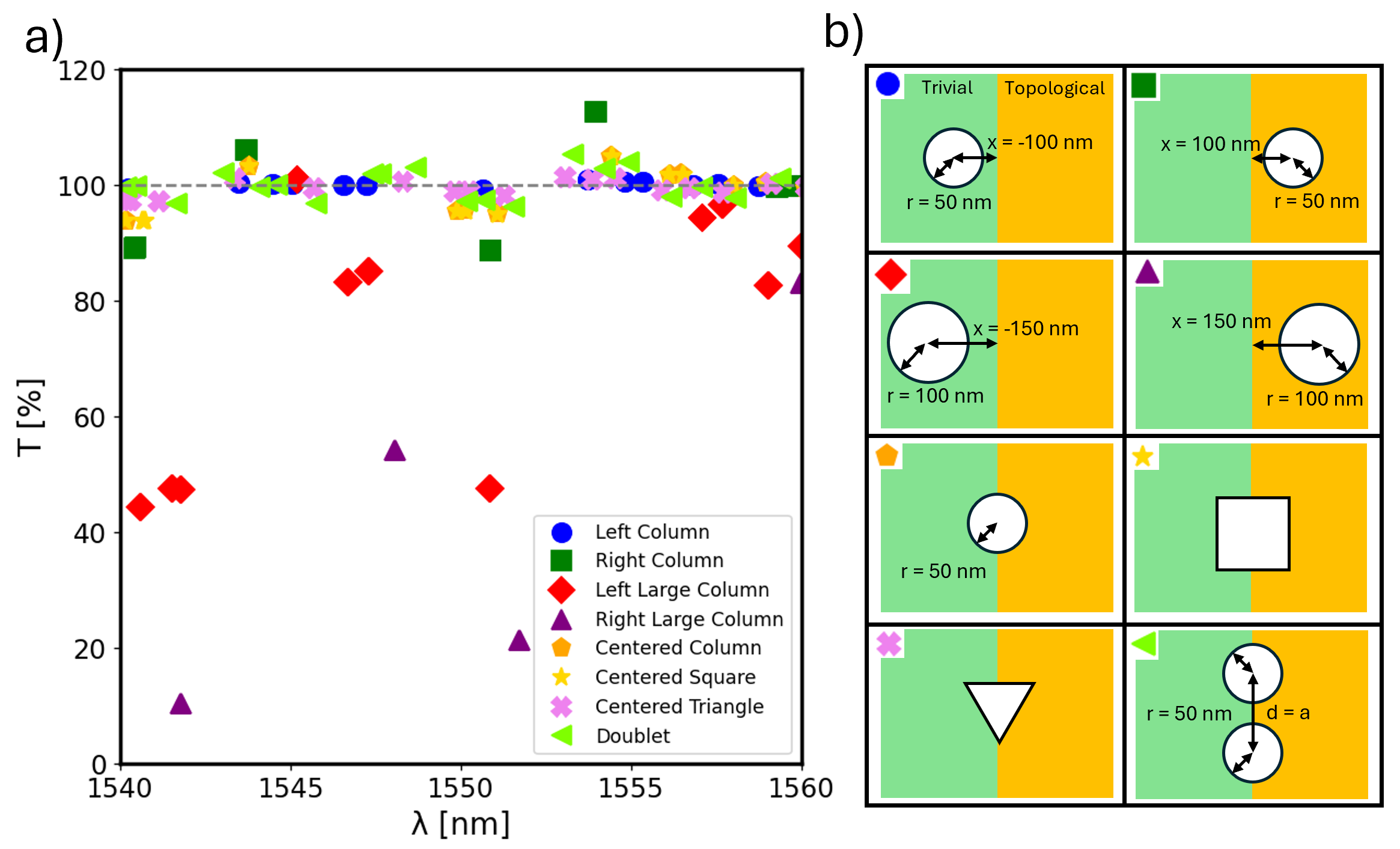}
\caption{Defect tolerance of the interface mode under eight localized perturbations.
(a) Plot of $T$[\%], defined as the ratio of transmitted field intensity through the defected interface relative to the defect-free case. 
(b) Geometries and placements of the defects, located near the modal maximum along a straight interface.}

\label{fig:defect_geometry}
\end{figure*}

To further assess the sensitivity of the mode to localized geometric perturbations, we conducted an additional series of simulations with eight different defect types, made out of the same material as both PCs, introduced along a straight interface. These included variations in shape (circle, square, triangle), size (50 or 100 nm radius), and placement (left or right side, or centered across the interface). The design and exact location of each defect are illustrated in Fig.~\ref{fig:defect_geometry}b. The numerical simulations of the near field revealed a strong asymmetrical localization of the mode around the interface: the modal maximum is slightly displaced toward the non-trivial side, while the field decays more slowly into the trivial side. To intentionally maximize the potential impact of each defect, we embedded them near the modal maximum and symmetrically around the interface, targeting regions where the field amplitude remained significant. This placement strategy was designed to probe the sensitivity of the mode to perturbations in both high- and moderate-field regions, thereby providing insight into the spatial asymmetry and extent of the mode profile.

The resulting transmission values across the topological band gap are plotted in Fig.~\ref{fig:defect_geometry}a. Most defects have little effect on the interface mode propagation, indicating good robustness. However, three cases exhibit a significant drop in transmission: the ``right column'' (a cylindrical defect with radius $r = 50$\,nm, displaced 100\,nm to the right side of the interface), the ``large right column'' (radius $r = 100$\,nm, displaced 150\,nm to the right), and the ``large left column'' (radius $r = 100$\,nm, displaced 150\,nm to the left). While the impact of the large defects is expected due to the substantial disturbance of the local geometry, the relatively strong suppression caused by the right column defect is less intuitive. We suspect that this behavior may be related to the spatial distribution of the edge mode: its amplitude is larger on the non-trivial (right-side) crystal, while its evanescent tail extends further into the trivial (left-side) region. As a result, the mode may be more sensitive to perturbations placed directly within the high-amplitude region (right side), whereas defects on the left produce a weaker but more spatially extended influence. This asymmetric sensitivity might also help explain the reduced transmission observed in sharp left turns, where overlapping tails within the bend region could interfere destructively. This point is further elaborated in \ref{ap_field_structure}.

These findings reinforce the interpretation that the interface mode arises from a polarization mismatch between OAL phases. While bulk-boundary correspondence guarantees the existence of interface-localized modes, their robustness depends on the local preservation of the symmetries that protect them, such as inversion or rotation. In our system, asymmetries in unit cell placement or the handedness of the interface can locally break these protections and reduce mode resilience. It should also be noted that the transmission values shown are obtained from FDTD simulations and exhibit numerical oscillations; they should therefore be understood qualitatively rather than quantitatively. Moreover, the plotted ``transmission'' represents the ratio of field intensity through a defected interface compared to an otherwise identical defect-free one and should thus be interpreted as a relative measure of mode suppression rather than an absolute transmission coefficient.

\subsection{Cell-count effect}
\label{cell_count_effect}
\noindent
An important yet often overlooked parameter in the design of topological photonic interfaces is the number of photonic crystal (PC) periods surrounding the interface. This parameter critically affects how well the interface-localized modes decouple from each other and localize spatially.

Because the interface in our design hosts two counter-propagating modes, arising from polarization mismatch across OAL phases, sufficient separation is needed to prevent overlap between the two interfaces within the supercell. To quantify the spatial localization and directionality of each mode, we divide the supercell into two equal regions centered on the left and right interfaces. We then integrate the field energy of the eigenmodes in each region to determine their confinement.

\begin{figure*}[t]
\centering
\includegraphics[width=\textwidth]{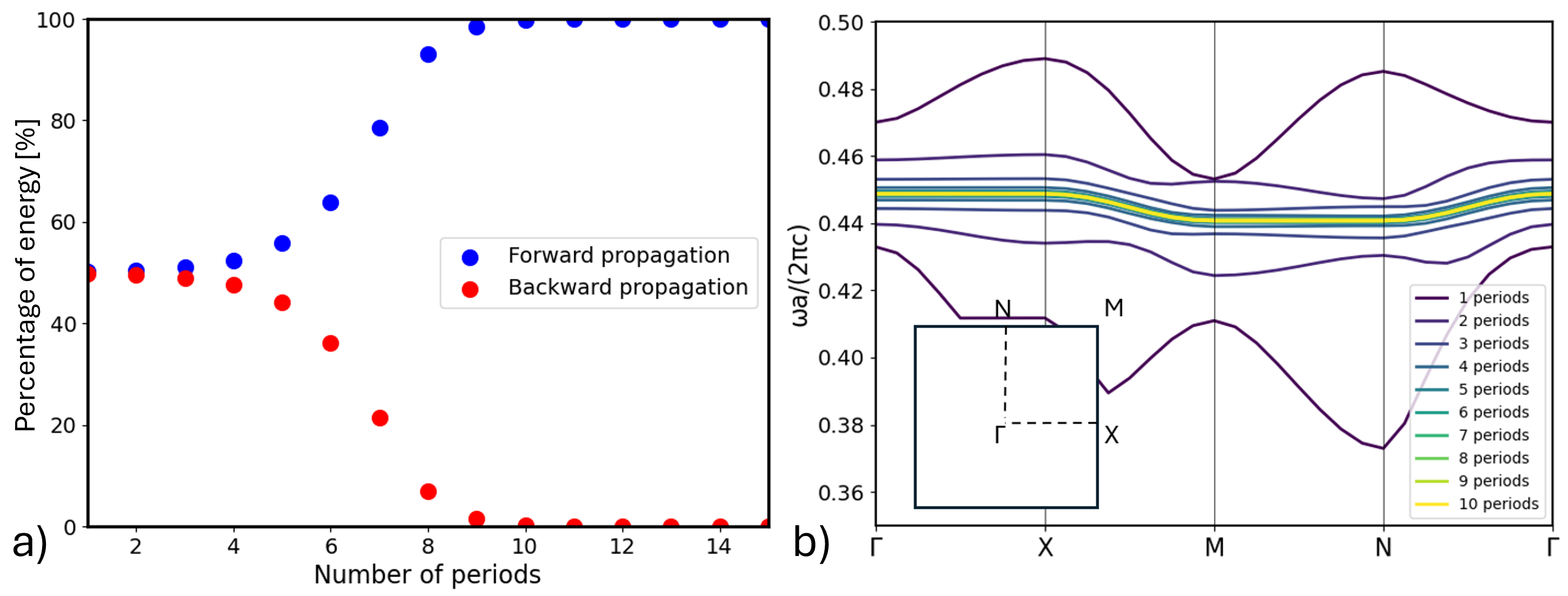}
\caption{
\textbf{Impact of unit cell count on mode localization and spectral stability.}
(a) Energy confinement of the interface mode to either the left or right side as a function of the number of PC periods. Values approach near-perfect confinement ($>$99\%) with 10 or more periods.
(b) Band structure evolution of the topological interface mode with increasing unit cell count. The dispersion converges as the mode becomes more localized.
}
\label{periods_effect}
\end{figure*}

\begin{figure}[t]
\centering
\includegraphics[width=0.7\textwidth]{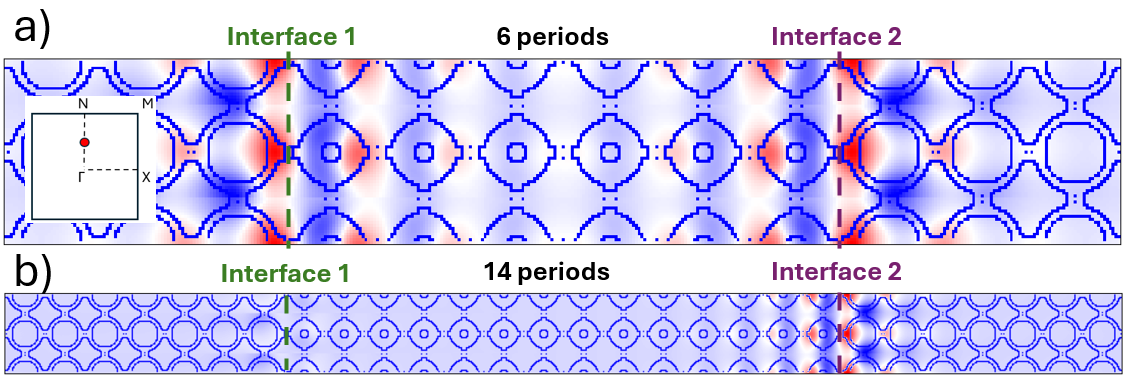}
\caption{
\textbf{Mode field distribution for different interface sizes.}
$E_z$ field profiles of the lower interface mode for (a) 6 periods and (b) 14 periods around the interface. The field becomes more localized with increasing interface width. 
Field shown along the $\Gamma$–$N$ direction; structural outlines are schematic and do not fully reflect fabrication detail. For accurate geometry, see Fig.~\ref{structure_scheme}c.
}
\label{field_distribution}
\end{figure}

As shown in Fig.~\ref{periods_effect}a, increasing the number of surrounding unit cells causes each mode to become more strongly confined to a single interface. At around 10 periods, the energy becomes almost fully localized to one side (over 99\%), enabling effectively unidirectional propagation. This is also evident in the field distributions: for a 6-period configuration (Fig.~\ref{field_distribution}a), the field still spreads toward both interfaces, whereas with 14 periods (Fig.~\ref{field_distribution}b), the mode is fully concentrated at the far interface.

This spatial separation also stabilizes the spectral position of the interface modes. As shown in Fig.~\ref{periods_effect}b, increasing the number of periods leads to convergence of the interface bands to a well-defined frequency range. This convergence ensures robust design targeting and minimal sensitivity to small changes in geometry or fabrication tolerances.

These results demonstrate that the number of periods around the interface is a key design parameter. It determines not only how well the modes are confined, but also whether true one-way routing and clean spectral isolation can be achieved, both essential for integrated photonic applications such as waveguide junctions and logic circuits.

\section{Conclusion}

\noindent
We have demonstrated a photonic crystal interface design that supports symmetry-protected edge modes arising from a bulk polarization mismatch between two distinct OAL phases. These modes exhibit characteristics commonly associated with topologically protected states, such as confinement to the interface, dispersion within the bandgap, and robust transmission, despite the absence of a nontrivial $\mathbb{Z}_2$ invariant. Our approach relies on simple geometric tuning of a single-material photonic crystal composed of InP, a well-established material compatible with standard lithographic techniques. The design is spectrally targeted to 1550\,nm, making it directly relevant for integrated photonic circuits in telecommunication applications.

Full-wave simulations reveal that the interface modes exhibit nearly lossless transmission along straight waveguides, and retain high performance across a variety of localized defects and geometric perturbations, particularly when these are introduced on the symmetric or “topological” side of the interface. However, left–right asymmetry in defect sensitivity highlights the partial nature of the protection, which is rooted in spatial symmetry and Wannier center displacement rather than global topological indices. We further explored how the number of photonic crystal periods surrounding the interface impacts mode localization and spectral stability. With 10 or more unit cells on each side, more than 99.7\% of the modal energy becomes confined to a single interface, allowing effective unidirectional propagation. This confinement not only enhances robustness but also stabilizes the interface band structure.

Altogether, these results establish OAL-based polarization contrast as a practical and tunable strategy for realizing robust edge transport in photonic systems, particularly where full topological protection is challenging to implement. Our findings open new pathways for scalable, lithography-compatible, and compact photonic devices leveraging bulk-interface polarization mismatch.

\section*{Acknowledgments}
We acknowledge financial support from Projects No. PID2023-152225NB-I00, and Severo Ochoa MATRANS42 (No. CEX2023-001263-S) of the Spanish Ministry of Science and Innovation (Grant No. MICIU/AEI/10.13039/501100011033 and FEDER, EU)"), Projects No. TED2021-129857B-I00 and PDC2023-145824-I00, funded by MCIN/AEI/10.13039/501100011033 and European Union NextGeneration EU/PRTR and by the Generalitat de Catalunya (2021 SGR 00445), We also acknowledge the financial support from the Charles University, project GA UK No. 80224.

\section*{Disclosures}
The authors declare no conflicts of interest.

\section*{Ethics statement}
This study did not involve human participants, human data or tissue, or animal subjects.

\section*{Data Availability Statement}

The data that support the findings of this study are available from the corresponding author upon reasonable request.

\bibliographystyle{bibft}\it
\bibliography{bibfile}

\appendix

\section{MPB calculations}\label{appMPB}

For more details on MPB please visit the documentation https://mpb.readthedocs.io/en/latest/

\subsection{Material model}
At this date, MPB is not capable of model materials with dispersion. Material parameters can be anisotropic and complex (as long as the eigenvalues of the complex representation remains real, which is related to lossless materials), but there is no dependency on frequency. 

For this reason, the calculated band structure is valid only for a region (y-axis) which corresponds to the wavelength at which the permittivity is defined. In our case, we use InP at 1550 nm, which corresponds to $\varepsilon_r = 10.01659$

This is approximation is valid as it holds in the region of the PC band-gap and hence the topological states. 

\subsection{Interface simulations}

\begin{figure}[t]
\centering
\includegraphics[width=0.48\textwidth]{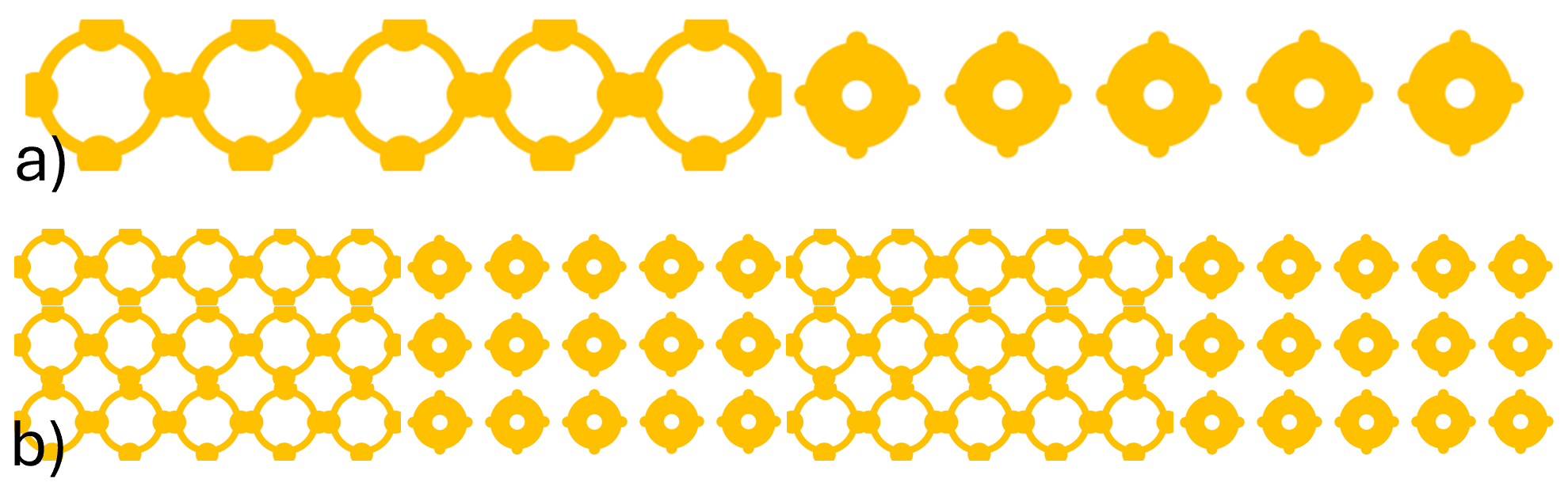}
\caption{Example of a interface construction in MPB. a) Elementary cell with 5 periods of each design, which will be periodically repeated. b) Combination of several elementary cells.}\label{interface_example}
\end{figure}

MPB can calculate only periodical systems, which means that defining an interface as a boundary between two semi-infinite planes is not possible. In our calculations, we use use several periods of the two complementary PC design as shown on figure \ref{interface_example}a, which produces a pattern shown on fig. \ref{interface_example}b. Using sufficient number of periods is required to simulate an interface between two PCs. In section \ref{cell_count_effect} we discussed, how the number of periods around the interface affects the formation of topological bands and their properties. It can also serve as a convergence test for the simulation validity. In ideal case, the number of periods N would be infinite, which would perfectly mimic the two semi-infinite PCs, but as we can see in figure \ref{periods_effect}, increasing N over 10 does not bring any noticeable change in properties of interest.

\section{FDTD simulations}
\label{appFDTD}

We perform our FDTD simulations using Ansys Lumerical FDTD solver.

\subsection{Material model}
Unlike MPB, FDTD material model supports dispersion so this effect is already taken into account in transmission simulations. In FDTD, materials are modelled by polynomials over the spectral region of interest with Kramers–Kronig relations in mind. The dielectric tensor dispersion of the material model in our simulations is up to sixth polynomial order with no special weights for either real or imaginary part. In the simulation spectral region, the imaginary part of permittivity of InP is already negligible and we treat it as if it would be zero. 

\subsection{Topological source}
In our simulations, we split the source preparation into two steps. In the first step, we used a combination of dipole sources with phase offset to evoke the topological modes in a straight interface and recorded the field at the other end of the interface. In the second step, we imported the recorded field as a new source in the simulation with the geometry that we want to investigate.

\subsection{Boundary conditions}
In contrast to MPB calculations, FDTD simulations do not use periodic boundary conditions. Instead, absorption layers are employed to prevent backscattering at the edges of the simulation region. This approach demonstrates that the results obtained from periodic MPB calculations are valid and translate well to finite systems.

\subsection{Numerical oscilations}
In figure \ref{transmission_maps}e, it is apparent the transmission reaches values higher then 100\% which is not physically correct. The problem is that the FDTD simulation operates in a time domain, which means that the spectral data are recorded by Fourier transformation. Since the length of the simulation is not infinite, the recorded data will be affected by oscillating error. This error is most apparent at the edges of the topological modes where the transmission sharply rises from 0 to 100\%. Furthermore, since we are interested in transmission as a fraction of output field to input field, the oscillation in denominator causes further noise in the result. This fraction error is apparent in \ref{transmission_maps}e left turn and defect left data. 
After strong consideration, we decided that editing the data would only hurt the results. Cutting it to 100\% would affect only the over-oscillations. Smoothing it would not help either, as some values might drastically overshoot and create a false impression of neighboring values being higher than they actually are. For this reason, we show this only as a qualitative demonstration rather then quantitative. 

\subsection{Field detail with underlying structure}
\label{ap_field_structure}

\begin{figure}[t]
\centering
\includegraphics[width=0.5\textwidth]{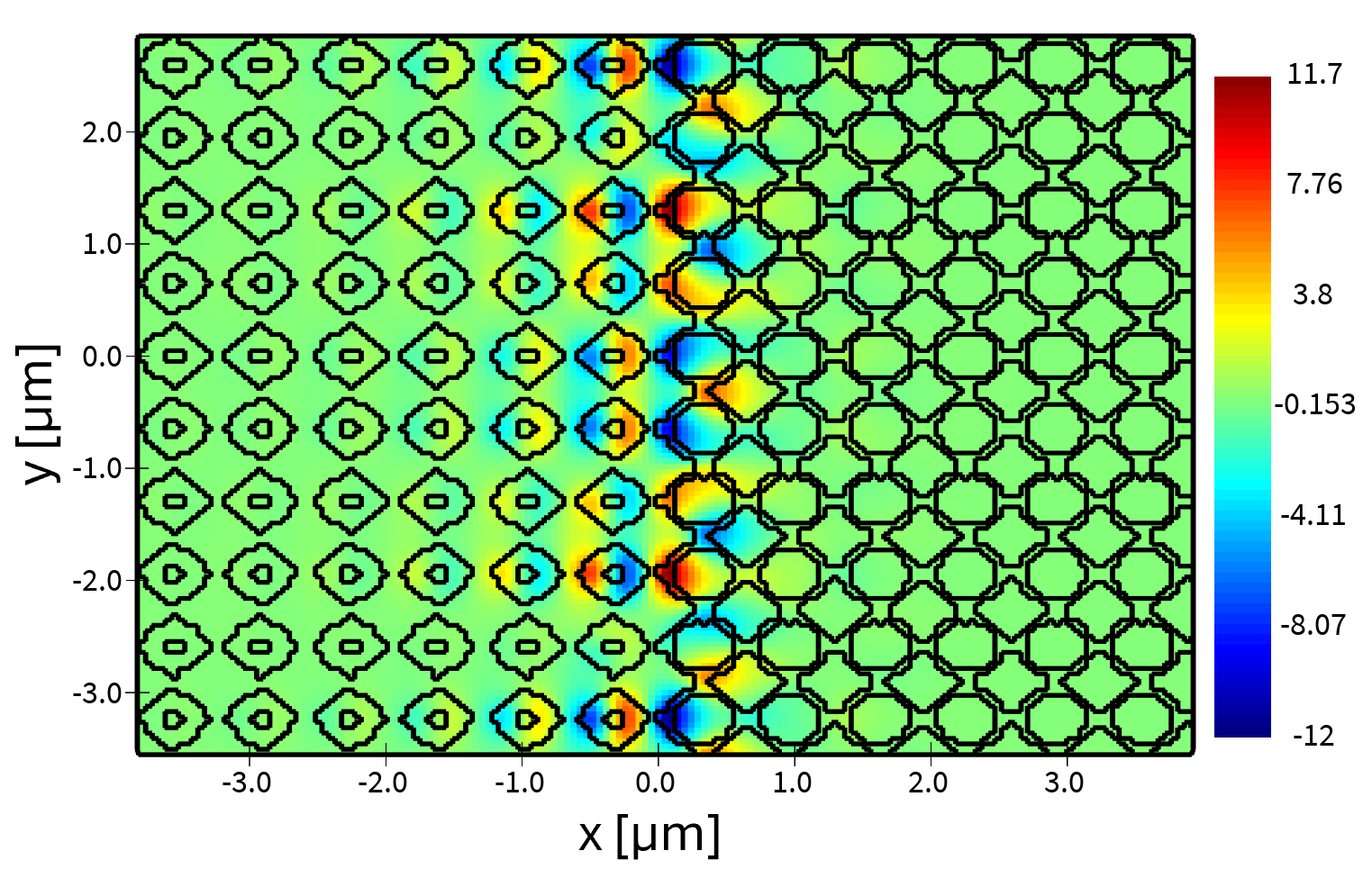}
\caption{
Detail of the $E_z$ field distribution at $\lambda=1550$nm overlaid with the underlying photonic crystal structure, obtained from FDTD simulations.}
\label{field_structure}
\end{figure}

In the main text, we noted that the topological mode exhibits stronger field lobes on the non-trivial side of the interface, while its evanescent tail extends further into the trivial side. This behavior is illustrated in Fig.~\ref{field_structure}, which shows the $E_z$ component of the electric field from the interface simulation (focused on a smaller region near the interface). The field profile reveals that the highest amplitudes and most pronounced lobes are localized within the non-trivial photonic crystal (right side), concentrated near the interface. Notably, the mode exhibits minimal penetration into the second column of non-trivial unit cells, indicating a sharp spatial confinement. In contrast, on the trivial photonic crystal side (left), the field lobes are significantly weaker, but the mode penetrates more deeply, reaching even the third column of unit cells. This asymmetry extends the spatial sensitivity of the mode on the trivial side and supports the interpretation of the mode as exhibiting directional field decay.

The staircased contours arise from the way Lumerical represents the underlying structure. For simulation purposes, staircasing, filling factor corrections, and smoothing are applied, which improve the convergence toward the actual geometry.

\end{document}